\documentclass[twocolumn]{aastex6}
\def\apj{ApJ}
\def\apjl{ApJ}
\def\apjs{ApJS}
\def\ao{Appl.~Opt.}
\def\aap{A\&A}
\def\mnras{MNRAS}
\def\prd{Phys.~Rev.~D}
\def\nat{Nature}

\def\et3{\eta_3}
\def\th1{\theta_{-1}}
\def\r07{r_{0,7}}
\def\x05{x_{0.5}}

\def\cm{\hbox{~cm}}
\def\kpc{\hbox{~kpc}}
\def\Mpc{\hbox{~Mpc}}
\def\Gpc{\hbox{~Gpc}}

\def\TeV{\hbox{~TeV}}
\def\GeV{\hbox{~GeV}}
\def\MeV{\hbox{~MeV}}

\def\G{\hbox{~G}}
\def\erg{\hbox{~erg}}

\begin{document}
\title{Properties of the Intergalactic Magnetic Field Constrained by Gamma-ray Observations of Gamma-Ray Bursts}
\author{P. Veres\altaffilmark{1}, C.~D. Dermer\altaffilmark{2},  K.~S. Dhuga}
\affiliation{Department of Physics, George Washington University, Washington, D.C. 20052 }
\altaffiltext{1}{Present address: Center for Space Plasma and Aeronomic Research, University of Alabama, Huntsville, AL  35899}
\altaffiltext{2}{Also, School of Physics, Astronomy, and Computational Sciences, George Mason University, Fairfax, VA 22030}

\begin{abstract}
The magnetic field in intergalactic space gives important information about
magnetogenesis in the early universe.  The properties of this field can be
probed by searching for radiation of secondary e$^+$e$^-$ pairs created by TeV
photons, that produce GeV range radiation by Compton-scattering cosmic
microwave background (CMB) photons. The arrival times of the GeV ``echo"
photons depend strongly on the magnetic field strength and coherence length.  A
Monte Carlo code that accurately treats pair creation is developed to simulate
the spectrum and time-dependence of the echo radiation.  The extrapolation of
the spectrum of powerful gamma-ray bursts (GRBs) like GRB~130427A to TeV
energies is used to demonstrate how the IGMF can be constrained if it falls in
the $10^{-21}$ -- $10^{-17}$ G range for a 1 Mpc coherence length.
\end{abstract}

\section{Introduction}
\label{sec:intro}
Magnetic fields are ubiquitous in cosmic sources ranging from stellar mass
objects to clusters of galaxies. Little information is known, however, about
the intergalactic magnetic field (IGMF) on the largest scales of the voids.
The properties of the IGMF, which are linked to cosmological structure
formation \citep{Neronov+09igmf},  result from processes in the early universe
or by expulsion of magnetic flux from structured regions. The characterization
of the IGMF is crucial to assess magnetogenesis and effects of structure
formation.  Multiple spectral \citep{2010Sci...328...73N}, angular
\citep[e.g.,][]{Dolag+09igmf,2010ApJ...722L..39A}, and temporal \citep{plaga95}
methods have been devised to constrain the magnitude of the average value of
the IGMF, $B_{\rm IGMF}$. Here we examine the temporal method involving delayed
echo emission from GRBs.

Although we confine our study to GRBs, the method is in principle applicable to
any flaring TeV source.  The scenario considered here consists of TeV range
source photons interacting with the extragalactic background light (EBL)
photons, creating e$^+$e$^-$ pairs.  The pairs lose energy by Compton
scattering CMB photons to the GeV range.  Because of the magnetic deflection of
the pairs, off-axis TeV photons generate GeV range $\gamma$-rays that travel to
the observer on longer path lengths resulting in a ``pair echo", delayed
compared to the prompt emission.  The delay time method was first presented in
\citet{plaga95} and later developed in papers by \citet{Razzaque+04gev},
\citet{Murase+08blazarigmf}, \citet{Ichiki+08igmf}, and
\citet{Takahashi+11highzGRB}.

The coherence length $R_{\rm coh}$ characterizing the distance over which the
IGMF changes direction by $\approx 90^\circ$, is a second important property of
the IGMF. Because the coherence length in intergalactic space is so poorly
known, the $\gamma$-ray techniques jointly constrain $B_{\rm IGMF}$ and
$R_{coh}$, rather than each individually. 

While there are no direct measurements of very high energy (VHE; E $\gtrsim$
0.1 TeV) radiation from GRBs, there are candidate events that under favorable
observing conditions might have produced a detection.  The spectrum of
GRB\,941017 \citep{Gonzalez+03PL} had a hard power law extending to $\gtrsim
100 \MeV$ with no turnover, in addition to the usual Band function describing
the MeV emission.  Similar hard power laws extending to multi-GeV energies have
been discovered by Fermi in the case of GRB\,090902B ($z = 1.822$)
\citep{Abdo+09-090902B}, GRB\,090926A ($z = 2.106$)
\citep{Ackermann+11-090926}, and GRB\,130427A ($z = 0.34$)
\citep{Ackermann+14_130427a}.  The direction towards GRB\,130427A was observed
with VERITAS \citep{Aliu+14130427a}, but the observing conditions were
unfavorable  and no VHE detection was made.  

For the redshift range $z\gtrsim0.3$, where most GRBs are detected, the pair
formation optical depth of TeV photons with EBL photons is $\gg1$, so VHE
emission from high-redshift GRBs would be strongly attenuated.  Indeed, the
highest energy GRB photons yet measured, for example, the 95 GeV photon from
GRB\,130427A measured a few minutes after the burst trigger, are detected from
relatively low redshift GRBs.  GRBs are therefore reasonable candidates for
TeV-range emission arising from either internal or external shocks, though we
must assume that the hard GeV component continues uninterrupted up to photon
energies $E\gtrsim 1 \TeV$.

In this paper, we present Monte Carlo simulations of the above mentioned
process for pair echo emission. The simulation assigns pair energies following
their proper distribution so the pairs don't necessarily take half of the TeV
photon's energy, a point often neglected in the literature.  We give detection
prospects for GRB echo radiation by the {\it Fermi} Large Area Telescope, by
existing air and water Cherenkov telescopes, and by the future Cherenkov
Telescope Array (CTA).  We employ a threefold approach to constrain the value
of the IGMF.  We examine the echo radiation observables for the extremely
bright GRB 130427A, and the detection prospects for Cherenkov telescopes at TeV
energies for this same GRB.  We also consider long exposure observations of
GRBs with hard high-energy spectral components.

Our calculations are made for a flat $\Lambda$CDM cosmology with
$\Omega_m=0.27$, $\Omega_{\Lambda}=0.73$, and H=72 km s$^{-1}$ Mpc$^{-1}$.  For
simplicity we use the terms pairs and electrons interchangeably.  For quantity
$Q$, we use the  $Q_x = Q/10^x$ scaling notation.  Physical constants have
their usual notation.  In Section 2 we review the analytic approach, and in
Section 3 we describe our Monte Carlo simulation. Results are presented in
Section 4, and we discuss and conclude in Section 5.

\section{Analytic considerations}
\subsection{VHE photons}

The source, in our case a GRB, emits TeV range ($0.1 \lesssim E=
E_{\rm TeV}$ TeV $\lesssim 100$ TeV)  photons which interact with EBL photons
on the pair-production distance scale
$\lambda_{\gamma\gamma}(E)$. This can be calculated for a specific EBL model
that provides an optical depth $\tau(E,z)$ at  photon energy E  and redshift
z, noting that  $\lambda_{ \gamma \gamma}(E) \cong D /\tau(E,z)$, 
where $D$ is the distance to
the source.

\begin{figure}
\begin{center}
\includegraphics[width=0.99\columnwidth,angle=0]{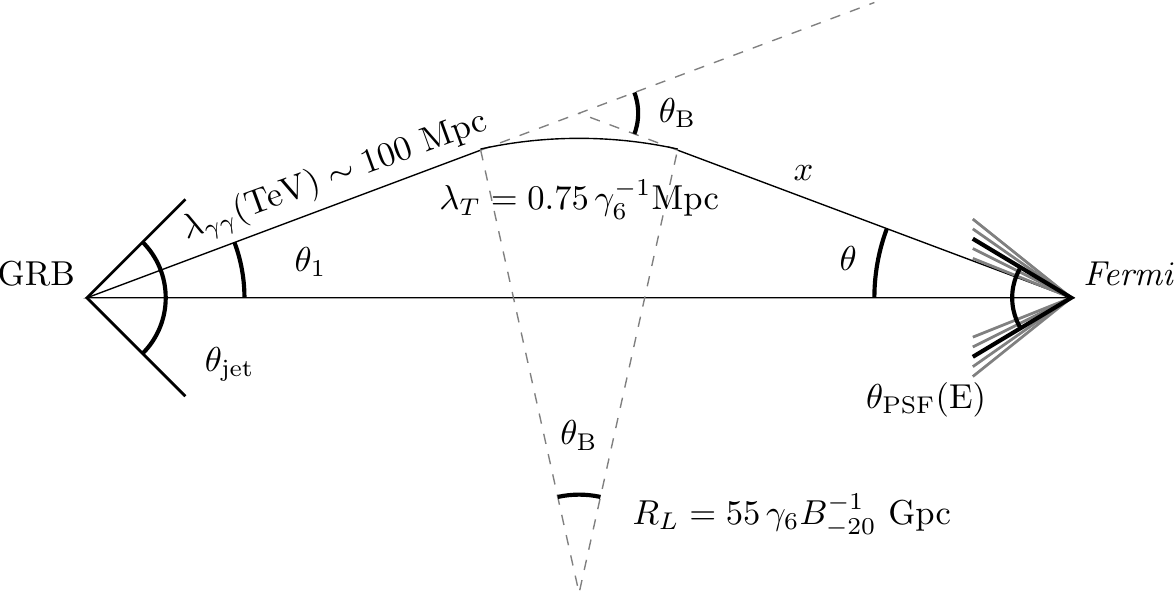}
\caption{Schematic diagram of the scattering geometry.  A photon from the GRB
is emitted at angle $\theta_1$ to the line of sight of an observer at distance
$D$ and pair produces by interacting with an EBL photon. The pairs with Larmor
radius $R_{\rm L}$ in a uniform field of strength $B$ cool on the length scale
$\lambda_{\rm T}$. Depending on the strength of the IGMF, the pairs may be
deflected before cooling and so are able to emit GeV photons in the direction
to the observer.  The array of observing angles indicate the energy dependence
of the point spread function (PSF) of Fermi/LAT. } \label{fig:cartoon}
\end{center} \end{figure}

This interaction yields an electron-positron pair, each with Lorentz factor
$\gamma_e = 10^6\gamma_6$ and energy $E_e = m_ec^2 \gamma_e$, that are {\it on
average} half of the energy of the TeV range photon, so that $\gamma_6\approx
E_{\rm TeV}$.  These pairs interact with the CMB photons and lose energy
through inverse Compton (IC) scattering.  The electron has a mean free path
$\lambda_e=(n_{\rm CMB} \sigma_T)^{-1}= 0.40 \kpc \, (1+z)^{-3}$ between
scatterings, where $n_{\rm CMB}\approx 409\, (1+z)^3 \cm^{-3}$ is the CMB
photon number density for temperature $T_{\rm CMB}=2.725~(1+z)~ {\rm K}$
\citep{Fixsen09TCMB}.  The electrons lose their energy through repeated
scatterings on a cooling length scale of 
\begin{equation}
\label{eq:ICcool}
\lambda_T(\gamma_e)=\frac{m_e c^2}{\frac{4}{3} \sigma_T u_{\rm CMB}
\gamma_e} = 0.72 ~(1+z)^{-4}~\gamma_6^{-1} \Mpc,
\end{equation}
where $u_{\rm CMB}\approx4.2\times 10^{-13} (1+z)^{4}\erg \cm^{-3}$ is the
energy density of the CMB. Electrons with $\gamma_6\approx1$ scatter on the
order of $\lambda_T/\lambda_e \sim 1000$ times before losing a substantial part
of their energy.  The upscattered photon's energy is $E_{IC} \approx 4/3~ (2.7~
k_B~ T_{\rm CMB}) \gamma_e^2 \cong 0.8 ~\gamma_6^2 \GeV $. The scattering is in
the Thomson regime provided $4\gamma_e(E/m_ec^2) \ll 1$, which implies that
$\gamma_e \ll 2\times 10^8$. Klein-Nishina effects therefore become important
only for $\gtrsim 100$ TeV photons and, though we use the full Klein-Nishina
kernel in our calculations, can be neglected in our study since the cutoff
energy of the GRB emission is  always assumed to be $\lesssim 30$ TeV.

The process is sketched in Figure \ref{fig:cartoon}.  
The angle between the direction of the VHE photon, and the
upscattered CMB photon is
\begin{equation}
\theta_B=\lambda_T/R_L=1.3\times 10^{-5}~ (1+z)^{-4}~ B_{-20}~ \gamma_6^{-2},
\end{equation}
where $R_L=\gamma m_e c^2/q_e B = 55 \gamma_6~B_{-20}^{-1} \Gpc$ is the Larmor
radius in a uniform field of strength $10^{-20} B_{-20}$ G.  This equation is
valid when the IGMF is coherent on scales larger than the IC cooling length
$\lambda_{T}$, given by Equation \ref{eq:ICcool}.  In the opposite case, the
deflection angle is modified by a factor of $\sqrt{R_{\rm coh}/\lambda_T}$,
reflecting the random walk of the electron as it cools.

Following the  notation of \citet{Dermer+11magn}, the time delay of the echo
photons arriving at the detector compared to the arrival time of photons that are 
emitted in the direction to the observer arrive can
be calculated from the differences in the path length, giving
\begin{equation}
\label{eq:dt}
c\Delta t = \lambda +x-D \approx \frac{1}{2}
\lambda_{\gamma\gamma} \theta_B^2
\left(1-\frac{\lambda_{\gamma\gamma}}{D}\right),
\end{equation}
where $\lambda = \lambda_{\gamma\gamma}+\lambda_{\rm T} \cong
\lambda_{\gamma\gamma}$, and we have expressed $x$ through the sine theorem:
$x=D \sin \theta_1/\sin\theta_B$ (see Figure \ref{fig:cartoon}).

Because the time delay can vary over many orders of magnitude depending on the
values of $B$, $R_{coh}$, $E$, and  $z$, it is instructive to give
order-of-magnitude estimates for the time delay in specific cases.  For $E=0.5
\TeV$ and $z=0.34$ (the redshift of GRB~130427A), we have
$\lambda_{\gamma\gamma}\approx 700 \Mpc $ and $\Delta t\cong 0.3B_{20}^{-2}$
year.  In the case of an $E=10 \TeV$ photon, the mean free path for pair
production with photons of the EBL is $\lambda_{\gamma\gamma}\approx 110 \Mpc $
and the time delay is  $\Delta t\approx 9 B_{-20}^{2}$ s.


\section{Monte Carlo simulation}

We have developed a code to calculate the echo flux from a source that emits
VHE radiation with a known spectrum. The code follows the interaction of the
photons with the EBL, the energy loss of the resulting pairs, and their
deflection by the intergalactic magnetic field. Subsequently the pairs scatter
CMB photons to produce emission in the energy range  of the Fermi-LAT.  We make
the calculation by generating a population of pairs from the assumed GRB
spectrum of TeV photons that interact with the EBL.  We numerically integrate
the IC contribution of pairs to obtain the desired observations. Instead of
using the MC method for IC scattering, we numerically integrate the single
electron IC emissivity to obtain the observed spectrum, because it is
computationally more convenient.

As a geometrical problem (see Figure \ref{fig:cartoon}), the distance $D$ to
the source and the values of $\lambda_{\gamma\gamma}$ and $\theta_{\rm dfl}$
uniquely describe the configuration. Constraints such as requiring the photon
emission angle to be within the opening angle the GRB jet ($\theta_1 <
\theta_{\rm jet}$), or requiring the arrival angle to be within the Fermi-LAT
PSF ($\theta < \theta_{\rm PSF} (E)$), are easy to apply. We assume that the
jet axis is pointed towards us \citep[see][for a study of off-axis
jets]{2010ApJ...719L.130N}.

\subsection{TeV range radiation}

We generate a distribution of photons in the specified VHE energy range assumed
to be described by a power-law distribution with high-energy cutoff.  To mimic
the EBL absorption and to select the population for pair creation we retain
photons for which a randomly generated number (uniformly distributed between 0
and 1) exceeds $1-e^{-\tau(E)}$.  The distance at which each photon pair
produces is drawn from a random exponential distribution with
$\lambda_{\gamma\gamma}(E)$ as the average.

\begin{figure}
\begin{center}
\includegraphics[width=0.99\columnwidth,angle=0]{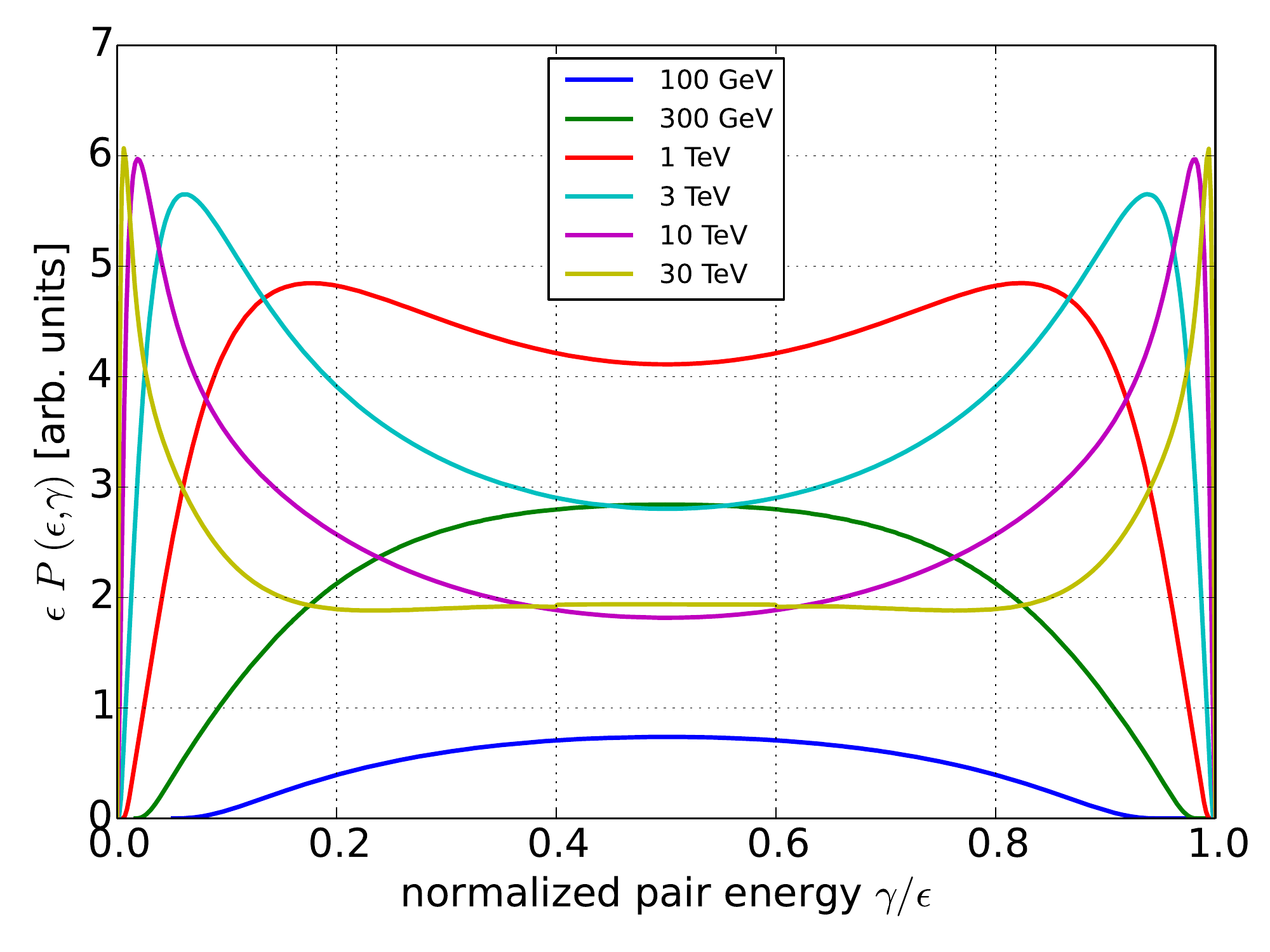} 
\caption{The normalized distribution of created pairs'
energy from VHE photons with different energies interacting with the CMB. }
\label{fig:zdz}
\end{center}
\end{figure}

The TeV photons produce pairs by interacting with the EBL. For analytical
calculations \citep[e.g.,][]{Dermer+11magn}  and some previous numerical
treatments \citep[e.g.,][]{Takahashi+11highzGRB, Fitoussi+17IGMF}, it is
customary to assume that the pairs equally share the energy of the parent TeV
photon. Here we use the appropriate distribution function for the pairs from
Equation (B1) in \citet{Z88pairs}.  The distribution in energies of an electron
with energy $\gamma_e$, given a VHE photon with energy $\epsilon m_ec^2 $, is
given by
\begin{widetext}
\begin{equation}
\label{eq:zdz}
P(\gamma_e,\epsilon)=\int\limits_{\frac{\epsilon}{4\gamma_e \gamma_e'}}^{\infty}
d\epsilon_{\rm EBL} n (\epsilon_{\rm EBL}) \frac{3 \sigma_{\rm T} c}{4 \epsilon^2
\gamma} \left( r-(2+r)\frac{\epsilon}{4\epsilon_{\rm EBL} \gamma_e \gamma_e'} +
2\left(\frac{\epsilon}{4\epsilon_{\rm EBL} \gamma_e \gamma_e'}\right)^2+
\frac{\epsilon}{2\epsilon_{\rm EBL} \gamma_e \gamma_e'}\ln \frac{4\epsilon_{\rm EBL}
\gamma_e \gamma_e'}{\epsilon}
\right)\;,
\end{equation}
\end{widetext}
where $\gamma_e'=\epsilon-\gamma_e$,
$r=(\gamma_e/\gamma_e'+\gamma_e'/\gamma_e)/2$, and $\epsilon_{\rm EBL}$ and
$n_{\rm EBL}$ are the EBL photon energy and number density, respectively,
provided in \citet{Finke+10ebl}. For different VHE photon energies, the
distribution is plotted in Figure \ref{fig:zdz}. The average value is indeed at
$\gamma_e=\epsilon/2$. At energies $\gtrsim \TeV$, however, the distribution of
secondary pairs  starts to become increasingly unequal.  For each VHE photon
with energy $E_{\rm TeV}=\epsilon\,$m$_e$c$^2$, we draw from this distribution
and retain $\gamma_e$ and $\gamma_e'$.

The pairs travel an average distance $\lambda_T$ before losing their energy by
scattering CMB photons. In our simulation, we again draw from an exponential
distribution with $\lambda_T(\gamma_e)$ as the mean. Only one generation of
secondaries is followed. The Compton-scattered radiation may again be
susceptible to pair attenuation by the CMB for the highest energy photons. For
the 30 TeV maximum energy of GRB photons assumed in this study, the second
generation emission in most of the cases can be neglected. The most energetic
echo photons come from the more energetic original photons, which interact with
the EBL, close to the source, namely at the same redshift. Where appropriate,
we apply EBL absorption to the echo spectrum. We use the model described in
\citep{Finke+10ebl} for the EBL,  which is similar to currently favored models
\citep{2015ApJ...812...60B,Stecker+16ebl}.

\subsection{Flux}

The differential distribution of pairs created from the interaction of TeV
photons with the EBL is denoted by $dN_0(\gamma_e)/d\gamma_e$. 
The echo radiation spectrum is calculated by integrating the electron
IC power over the electron distribution using the expression
\begin{equation}
\frac{d^2N_{\rm echo}}{d E_{\rm IC} d t}  = \int d\gamma_e \frac{dN
(\gamma_e)}{d\gamma_e} \frac{d^2 N_{E,{\rm IC}} }{dE dt}\,.
\end{equation}
The single electron power when scattering CMB photons is
\begin{equation}
\label{eq:ICpower}
\frac{d^2 N_{E, {\rm IC}} }{dE dt}= \frac{3\sigma_T
c}{4\gamma_e^2} \int \frac{d E_{\rm CMB}}{E_{\rm CMB}} n_{\rm E, CMB} f(x),
\end{equation}
where $f(x)= 2x \ln x + x + 1 -2x^2$  and $x= \epsilon m_e c^2/4\gamma_e^2 E_{\rm
CMB}$
\citep{Blumenthal+70IC}.
The time integrated IC photon flux can be obtained by multiplying Eq.
(\ref{eq:ICpower}) with the
local IC cooling time of the pairs: 
\begin{equation}
\Delta t'_{IC} (\gamma_e) = \frac{m_e c^2 }{\frac{4}{3} \gamma_e \sigma_T c~
u_{\rm CMB}} = 7.3\times 10^{13} ~ \gamma_6^{-1} ~ {\rm s}
\end{equation}
\citep{Fan+04TeV}.

In the analytic treatment of this problem, we need to link the distribution of
the electrons ($dN_0/d\gamma_e$) to the pairs that contribute within the
observing window ($dN/d\gamma_e$) \citep[e.g. ][]{dailu02, Dai+02igmf,
Razzaque+04gev}.  To accomplish this, one has to consider the maximum timescale
$\Delta t_{\rm obs} (\gamma)$ of the angular, magnetic-deflection, IC-cooling,
and GRB timescales as a function of $\gamma_e$ \citep{Razzaque+04gev}, giving
$dN/d\gamma_e = (\Delta t'_{IC}/\Delta t_{\rm obs} (\gamma)) dN_0/\gamma_e$
\citep{Dai+02igmf}.  In a numerical treatment of this process
\citep{Takahashi+11highzGRB}, the integration is performed between the
locations of pair production and energy loss, provided that the deflection
angle is sufficient for radiating emission into the observer's direction.

By contrast, in the MC approach used here we start from the distribution of
pairs generated using  Equation (\ref{eq:zdz}) and shown in Figure
\ref{fig:zdz}.  We select those individual electrons whose associated delay
time matches the observational criteria, and then calculate their contribution
to the echo radiation. 

We assume the source emits photons with a power law spectrum up to $E_{\rm
cut}$, which we typically choose to be 3 or 30 TeV in our examples. Photons
from this spectrum are absorbed by the EBL. We calculate the isotropic
equivalent energy absorbed from the difference between the unabsorbed and the
absorbed fluxes (as shown in the numerical calculations below).

\begin{equation}
{\cal E}_{\rm abs} = \int\limits_{10~{\rm GeV}}^{E_{\rm cut}} E \frac{dN(E)}{dE}
\left( 1-e^{-\tau (E,z)}\right) dE.
\end{equation}
The lower limit on the integration is set because the universe is transparent
to 10 GeV photons at all redshifts since the epoch of galaxy formation ($z \sim
10$).  The  number of pairs involved in our simulations is $\mathcal{O}(10^5)$,
but varies for different calculations.  We use the total absorbed energy ${\cal
E}_{\rm abs} $ to scale our simulation to the actual differential distribution
of pairs.  The simulated differential pair distribution is related to the real
distribution through a scale factor $C$ from the expression 
\begin{equation}
C \int\limits_1^{\infty} d\gamma_e
(dN_{\gamma_e} /d \gamma_e)_{\rm sim} \gamma_e m_e c^2= 
C \sum\limits_{i=1}^{N_{pairs}} \gamma_i m_e c^2 = {\cal E}_{\rm abs}.
\end{equation}

Next, we apply the observational criteria specifying start and stop times of
observations.  We choose the number of TeV photons, the cut energy $E_{\rm cut}$,
$z$, $B_{\rm IGMF}$, and $R_{\rm coh}$. We calculate $\lambda_{\gamma\gamma}$
and  $\lambda_T$. Using these values, we determine $\theta_{\rm dfl}$ and the
time delay $\Delta t$.  We calculate the individual electrons' contribution to the
spectrum and sum. Although we follow individual electrons, the resulting spectrum 
will be smooth, because we convolve the electron distribution with the Compton kernel.

\subsection{Geometry}

The observing angle $\theta$ of IC photons with energy $E_{\rm GeV}$ GeV that
reach the observer can be calculated from the relation $\sin \theta =
(\lambda_{\gamma\gamma}/D) \sin \theta_{\rm dfl}$. In order to be observed by
Fermi, this angle has to be less than the energy dependent PSF of the LAT.  For
this, we use the expression $\theta_{\rm PSF}\approx 1.68^\circ E_{\rm
GeV}^{-0.77} + 0.2^\circ e^{-10~ E_{\rm GeV}^{-1}}$, which is an analytical
approximation for the 95 \% containment PSF that is sufficiently accuate for
our purposes \citep[see][ for a detailed discussion  of the Fermi
PSF]{Ack+13psf}.  This constraint is important for large IGMF values $\gtrsim
10^{-17}$ G.  Neglecting this effect can introduce inaccuracies into the
low-energy part of the echo radiation spectrum.

\section{Comparison with observations}

There is a variety of available data and possible future observing scenarios
that can be useful in constraining the IGMF. Since we require VHE photons from
a transient source, the best candidates are GRBs with hard power law spectral
components in addition to the usual sub-MeV prompt spectrum.

Fermi LAT is an all-sky instrument \citep{Atwood+09LAT} sensitive in the 30 MeV
-- 300 GeV range.  Currently VERITAS \citep{veritas05}, HESS
\citep{Aharonian+04crab}, and  MAGIC \citep{Aleksic+12MAGIC} are the main air
Cherenkov observatories with pointing capabilities.  HAWC \citep{Ab+13HAWCsens}
is the largest operating water Cherenkov instrument that, though it does not
allow pointing, has  a large field of view and a nearly 100\%  duty cycle
compared to the $\sim 10$\% duty cycle of current imaging air Cherenkov
telescopes.  The Cherenkov Telescope Array, \citep[CTA][]{Actis+11CTA}, will
usher in a new era in  VHE astrophysics by having a sensitivity approximately
an order of magnitude greater than current air Cerenkov telescopes.

\subsection{Spectral and temporal evolution}
\begin{figure}
\begin{center}
\includegraphics[width=0.99\columnwidth,angle=0]{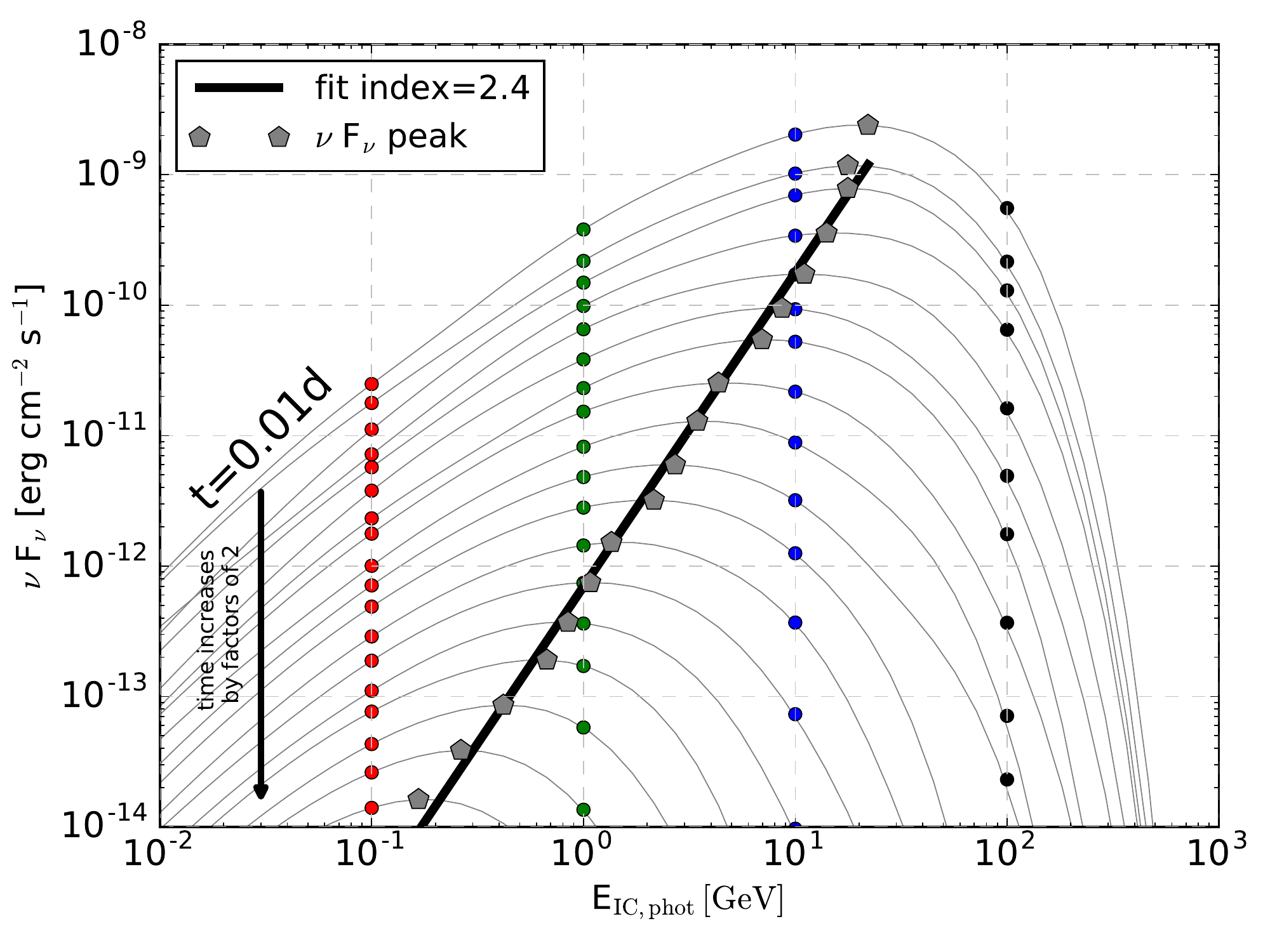} 
\includegraphics[width=0.99\columnwidth,angle=0]{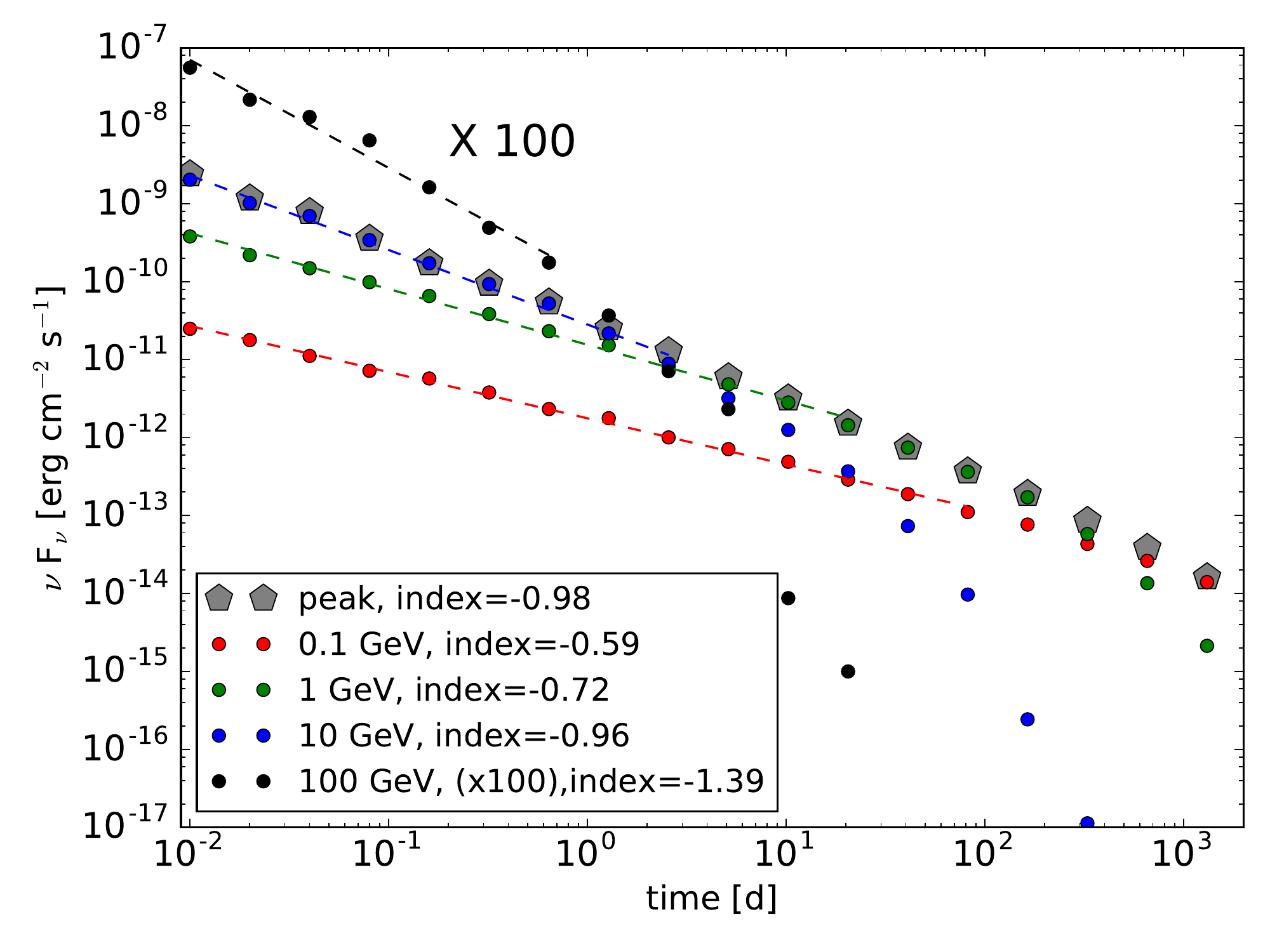} 
\caption{Time evolution of the echo radiation for $B=10^{-20}$ G and $R_{\rm
coh} = 1$ Mpc. The source of VHE photons (the ``C" interval of GRB 130427A, see
Figure \ref{fig:130427ap1}) is ommitted for clarity. The top figure shows echo
spectra recorded at 0.01, 0.02, 0.04 ... days after the trigger, with the lowest
curve at 1310 days. Gray pentagons
mark the maximum of the spectra, ($\nu F_\nu)_{\rm peak}\propto E_{\rm IC,
peak}^{2.4} $. The bottom figure shows the lightcurve at different energies and
at the peak, where we have ($\nu F_\nu)_{\rm peak}\propto t^{-0.98} $.}
\label{fig:b2}
\end{center}
\end{figure}

To assess if our code behaves correctly, we calculate the time evolution of the
echo flux for a cutoff of 3 TeV. Fig. \ref{fig:b2} shows results of our
calculations for $B = 10^{-20}$ G.  Here and throughout we fix the coherence
length of the magnetic field to $R_{\rm coh} = 1$ Mpc, while recognizing that
this is poorly known quantity. Derivations with different $R_{\rm coh}$ can be
calculated, however, since our data consists of upper limits, we focus this
study on the IGMF for an assumed value of $R_{\rm coh}$ to demonstrate the
method.

The spectra at different times have a low-energy power-law segment followed by
a break and a rapidly dropping flux (Fig. \ref{fig:b2}, top).  The echo
spectrum at low energies behaves roughly as $\nu F_\nu \propto E^{1.5}$,
corresponding to a cooling spectrum of a narrow electron distribution injected
at high energies.  The evolution of the break with time is a consequence of the
more rapid reprocessing of higher energy photons, which make higher energy
pairs that are less deflected by the IGMF, so that the higher-energy
Compton-scattered photons have a shorter path length to the observer.

Fig. \ref{fig:b2} (bottom) shows the passing of the spectral peak in different
energy bands.  At lower energies the power law slope is shallower and the break
occurs later.  The peak of the $\nu F_\nu$ spectrum decreases $\propto t^{-1}$.
This behavior broadly follows from the fact that the total echo fluence (F
$\times$ t) is determined by the amount of VHE flux absorbed by the EBL and is
constant.

\subsection{GRB~130427A and VERITAS constraints}
GRB~130427A had the largest $\gamma$-ray fluence of any GRB yet observed, with
high-energy emission detectable up to $\sim$1 day after the burst trigger with
Fermi LAT.  VERITAS followed up and derived an upper limit for the flux at 100
GeV at 0.82 days after the trigger \citep{Aliu+14130427a}.  Here, we use the
VERITAS upper limit to place constraints on the IGMF. We use the emission
episode from 11.5 -- 33 s after the burst trigger, where the spectrum can be
described as a power law with photon index $\Gamma=-1.66\pm 0.13$, to define
the $\gamma$-ray spectrum from this GRB.  Since there was no sign of a spectral
cutoff, we assume the spectrum extended up to 3 and 30 TeV to give estimates
for the echo radiation flux.

\begin{figure}
\begin{center}
\includegraphics[width=0.99\columnwidth,angle=0]{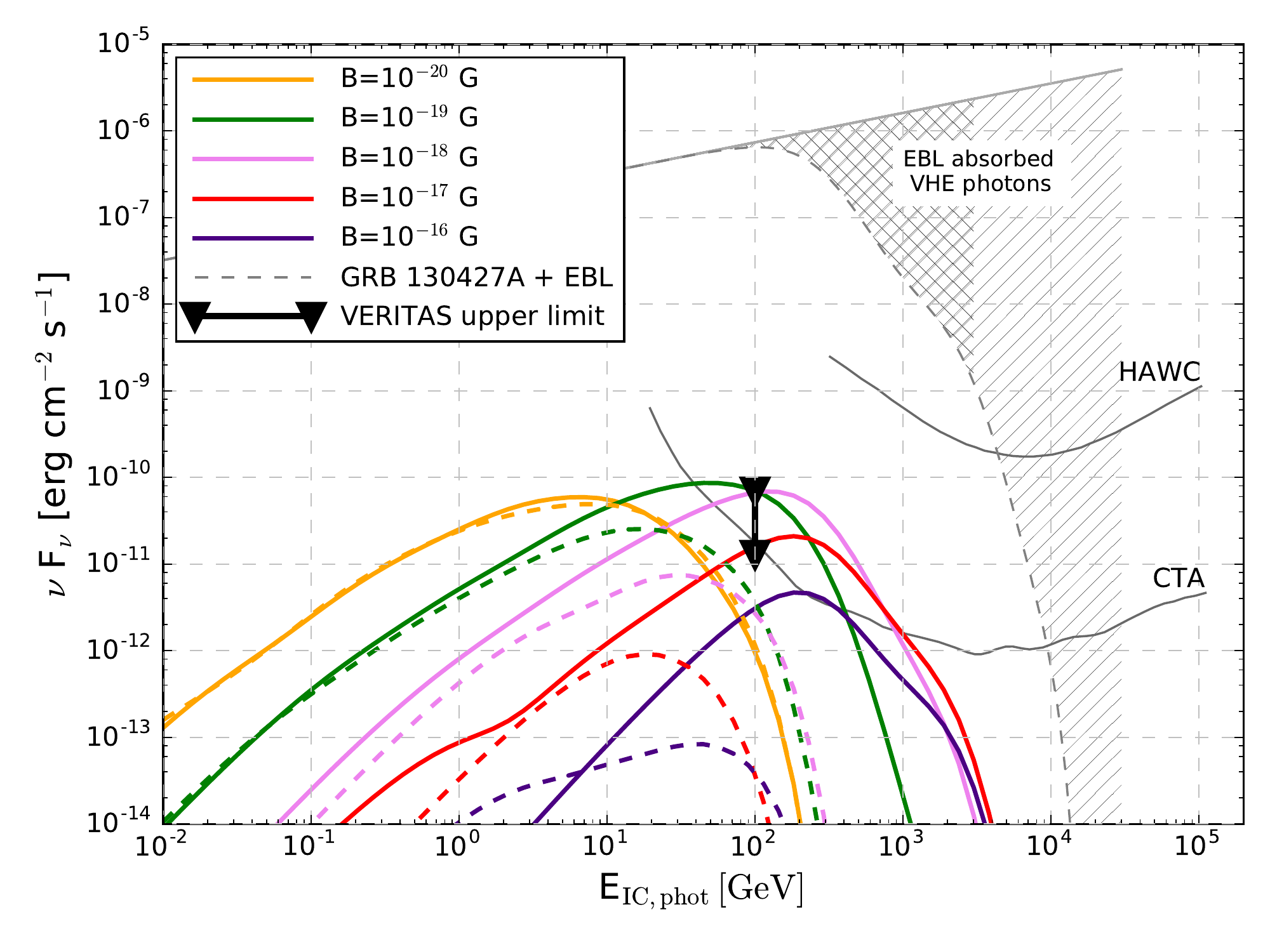} 
\caption{Simulated echo spectra assuming that the echo radiation is seeded by
an extrapolation of the GRB emission spectrum measured in interval ``C" ($\Delta
t=21.5$ s) of the lightcurve of GRB~130427A \citep{Ackermann+14_130427a} to TeV
energies. Numerical results with assumed cutoff energies of 3 and 30 TeV are
shown by the heavy dashed and solid curves, respectively, are compared with
VERITAS upper limit for GRB~130427A. The range of upper limits indicates
different assumed spectral index.  The coherence length of the IGMF is assumed
to be $R_{\rm coh}= 1\Mpc$. }
\label{fig:130427ap1}
\end{center}
\end{figure}

In Figure \ref{fig:130427ap1} we calculate the echo radiation using our
simulation and compare it with the VERITAS measurements. VERITAS provides a
range of upper limits based on the assumed spectral shape.  By comparing the
VERITAS limits with the simulated spectra, we find that either $ B_{\rm
IGMF}\gtrsim 10^{-17}$ G or $ B_{\rm IGMF}\lesssim 3\times10^{-19}$ G for the
$E_{\rm cut}=30 \TeV$ case (otherwise the calculated echo flux would violate
the VERITAS upper limits). The $E_{\rm cut}=3 \TeV$ case does not give any
constraints. Based on current estimates for CTA sensitivity, more stringent
constraints are expected for observations similar to that of VERITAS (see
Figure \ref{fig:130427ap1}). HAWC is more likely to observe the direct prompt
radiation than the echo flux.

\subsection{GRB~130427A  and Fermi-LAT constraints}
Another intriguing method to constrain the IGMF is provided by Fermi-LAT
observations of GRB~130427A.  The  $E=32 \GeV$ photon  observed at $\Delta t=
34.4$ ks after the GRB trigger is difficult to interpret  in the framework of
synchrotron radiation from the forward shock \citep{Ackermann+14_130427a}.
Nonetheless it might be associated with an IC component, even though there is
no evidence for the spectral and temporal break expected for a transition from
synchrotron to Compton emission.  

If we assume this photon originates from the echo
radiation, there is a straightforward way to estimate the IGMF using Equation
(\ref{eq:dt}). A simple calculation yields $B_{\rm IGMF}=2.2\times 10^{-19}
(\Delta t/34.4~ {\rm ks})^{1/2} (\lambda_{\gamma\gamma}/146~ {\rm Mpc})^{-1/2}
((1+z)/1.34)^{4} (E/32\GeV) \G$ with
$\lambda_{\gamma\gamma}\approx150\Mpc$ and  $E\sim 6.3 \TeV$ as the energy of
the primary photon. Equation (\ref{eq:dt}) can be solved by Monte Carlo methods
to gauge the error on this IGMF value, and we get $\log_{10} B_{\rm IGMF} =
-18.8\pm0.3$. This represents the average of the simulated values and it
differs somewhat from the estimated value, partly due to the asymmetry of the
distribution of simulated values.

\subsection{Long temporal baseline observations}
\begin{figure}
\begin{center}
\includegraphics[width=0.99\columnwidth,angle=0]{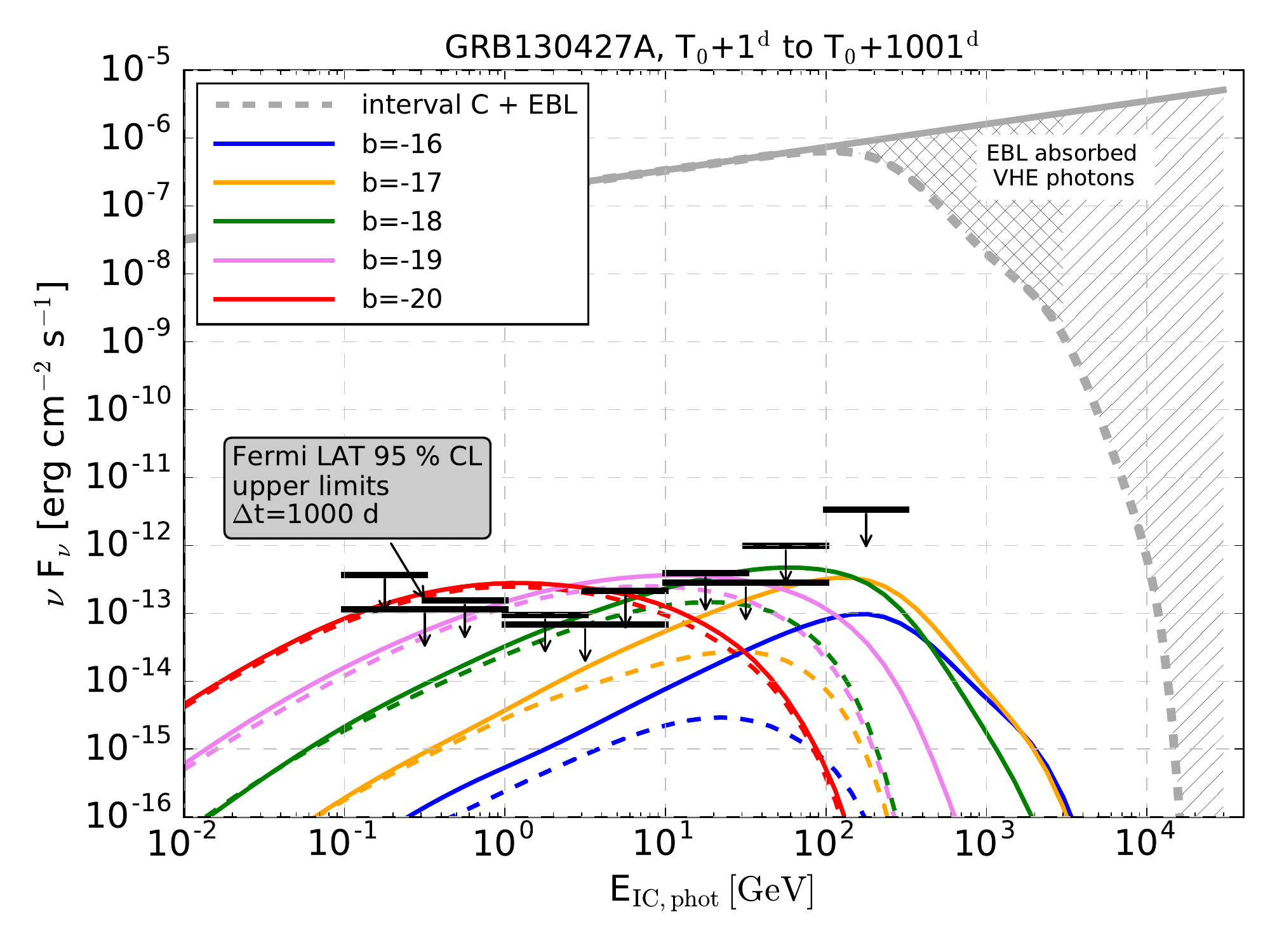} 
\caption{Long exposure observations with Fermi-LAT 
in the direction of GRB 130427A give the upper limits shown
for detection between 1 and 1001 days after the GRB as a function 
of photon energy for different values of $b$, where the IGMF $B = 10^b$ G. Calculations show 
echo spectrum for cutoff energies of 3 and 30 TeV vs. photon energy,
represented by dashed and
continuous lines, respectively. }
\label{fig:x4}
\end{center}
\end{figure}

The delayed echo radiation potentially lasts for a long time compared to the
time during which high-energy GRB afterglow radiation is expected. Since
Fermi-LAT is an all sky monitor, we explore the possibility that the  echo
radiation can be detected with long exposure observations starting from one day
after the GRB to several years, on the order of the lifetime of Fermi/LAT.
We therefore calculated the flux of the echo radiation from GRB 130427A for
observations starting one day after the GRB, where the direct  GeV range
afterglow radiation already went undetected.  This observation spans 1000 days
(see Figure \ref{fig:x4}). 

To compare the simulated echo spectrum with observations, we use Fermi-LAT
data, specifically PASS 8 photons with P8R2\_SOURCE\_V6 filters. The diffuse
galactic foreground was accounted for using gll\_iem\_v6, and the isotropic
diffuse radiation by iso\_P8R2\_SOURCE\_V6.  We used the {\tt
fermipy}\footnote{\url{http://fermipy.github.io} (version 0.13.2)} package and
its routines to derive Fermi-LAT upper limits.

Fermi LAT upper limits for the long term (1000 d) observations in the direction
of GRB~130427A are on the order of $10^{-13}$ erg cm$^{-2}$ s$^{-1}$ (see
Figure \ref{fig:x4} for the exact values). By comparing the upper limits to the
MC simulations, we can rule out the cases where $10^{-18} \lesssim B_{\rm IGMF}
({\rm G} )\lesssim 10^{-20}$. Stronger magnetic fields (B$_{\rm IGMF}\gtrsim
10^{-17}$ G) do not violate the upper limits.

\subsection{Cutoff energy}
\begin{figure}
\begin{center}
\includegraphics[width=0.99\columnwidth,angle=0]{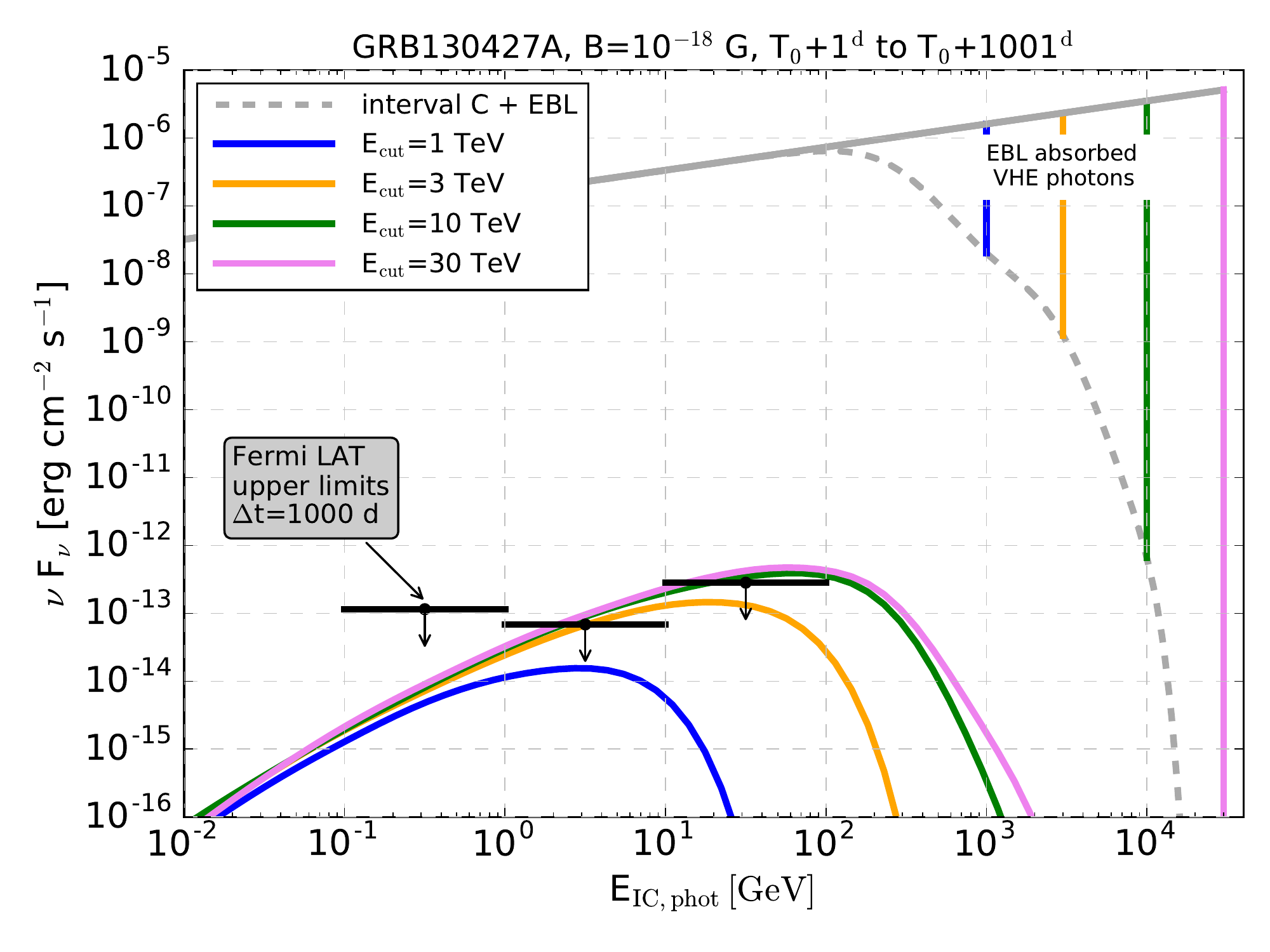}
\caption{Calculation to determine the cutoff energy $E_{\rm cut}$ for a GRB similar to GRB~130427A, 
assuming $B = 10^{-18}$ G and $R_{\rm coh} = 1$ Mpc.
Different values of $E_{\rm cut}$ are shown in the legend and the corresponding
echo spectrum is drawn with the same color. }
\label{fig:u1}
\end{center}
\end{figure}

The power-law spectral components with photon index harder than $-2$ will have
a cutoff energy.  This cutoff energy of the delayed hard component is generally
outside of the Fermi LAT range. To gauge the effects of different cutoff
energies, here we use Fermi LAT observations starting from one day after the
GRB trigger, where the afterglow has faded, ending at 1001 days after
(similarly to the previous subsection).  In Figure\ \ref{fig:u1} we make a
calculation of the echo radiation using the parameters of GRB~130427A, with
varying cutoff energies, assuming a magnetic field strength of $10^{-18}$ G. This is near
the lower limit of the IGMF inferred from observations of blazars
\citep{Dermer+11magn}. We overplot  the expected echo radiation with different
cutoff energies and show that long-term LAT observations provide meaningful
constraints assuming this value of $B$. In particular for $B= 10^{-18}$ G,
$E_{\rm cut}\gtrsim 3\TeV$  the predicted echo flux violates the LAT upper
limits.

\section{Discussion and Conclusion}

We have investigated the use of GRBs for constraining the IGMF using the
Fermi-LAT and existing and future Cherenkov telescopes. We assume GRBs emit
radiation in the TeV range, with pairs produced through $\gamma\gamma$ pair
creation of the VHE $\gamma$ rays interacting with EBL photons. These pairs
Compton-scatter EBL photons (here, we use the model of  \cite{Finke+10ebl}) to
GeV energies to make a delayed echo radiation.  We find Fermi LAT can constrain
the radiation from powerful GRBs like GRB~130427A if the IGMF is in the
$10^{-21}$ -- $10^{-17}$ G range, assuming a 1 Mpc coherence length.  Depending
on the assumed spectrum and cutoff energy of the TeV spectral component, this
range could be broader.  We have also shown how the VERITAS non-detection of
this GRB can constrain $B$ and the cutoff energy of the TeV spectrum.

Spectral methods to constrain the IGMF have been successfully applied in
studies of blazars \citep{2007A&A...469..857D, Neronov+09igmf,
2011ApJ...727L...4D, Dermer+11magn,Finke+15igmf} for which the VHE spectrum can
be directly measured.  Making the most conservative assumption that the blazar
is operating for no longer than the time over which high-energy $\gamma$-ray
emission has been observed, values of $B\gtrsim 10^{-19}$ -- $10^{-18}$ G are
inferred for a 1 Mpc coherence length of the IGMF.  This lower limit of the
IGMF is in the regime where pair echo radiation from GRBs can be detected, and
we have shown that echo radiation should be detected if the cutoff energy is
$\gg 1$ TeV for a GRB~130427A-type GRB. 

The inferences of the values of the IGMF from blazar studies is, like the GRB
case, very sensitive to the spectrum at the highest energies which cannot be
detected either due to sensitivity limitations or attenuation by the EBL.
Thus \citet{Arlen+14igmfsim} conclude that the blazar data is compatible with
an arbitrarily weak IGMF. 

The main advantage of GRBs for the temporal method of inferring the IGMF lies
in the transient nature of the direct emission. In the case of blazars, the
echo radiation is superposed on the direct emission, whereas for GRBs the
prompt emission fades away. The fading afterglow in the GeV range may still,
however, be confused with the echo radiation. \citet{Murase+09GRBEBL} discusses
the possible confusion between the echo and direct emission.  A further
pertinent issue for the inference of the IGMF from blazar studies is whether
collective effects from the beamed pairs extract the energy of the e$^+$ and
e$^-$ more quickly than IC processes \citep[e.g.,][]{2012ApJ...752...22B,
2012ApJ...752...23C,  2012ApJ...758..102S, 2015MNRAS.448.3405M}.  Depending on
the relative densities of the beam and background plasma, the duration of the
source, and pair spectrum, this energy can be extracted due to linear
two-stream instabilities. This issue, which is important for persistent sources
like blazars, is not so severe for a transient GRB source.

Fermi-LAT is the best suited instrument to probe the GeV range data produced by
the pair echo.  For a wide range of parameters likely to apply to this problem
(e.g. the value of $B$, the TeV-range spectrum of the GRB emission, and the
coherence length),  the peak of the echo spectrum falls in the Fermi-LAT range.
With the advent of HAWC and the upcoming CTA, searches for echo emission from
GRBs will provide more stringent constraints on the IGMF. 

{\bf Acknowledgments:} We thank Matthew Wood for prompt help using {\tt
fermipy}. PV acknowledges support from Fermi grant NNM11AA01A and thanks OTKA
NN 111016 grant for partial support.

\end{document}